\documentclass[twoside]{article}
\usepackage{graphicx}
\usepackage{fancyhdr}
\usepackage{multicol}
\usepackage{styl-ap}

\begin{document}
\title{On the 2011 outburst of the Recurrent Nova T Pyxidis}
\author{Luca Izzo\work{1}, Massimo della Valle\work{2}, Alessandro Ederoclite\work{3}, Martin Henze\work{4}}
\workplace{Dip. di Fisica, Sapienza University of Rome, P.le A. Moro 2, Rome; ICRANet, P.le della Repubblica 10, Pescara, Italy
\next
INAF - Osservatorio di Capodimonte, Salita Moiariello 16, 80131 Napoli, Italy
\next
Centro de Estudios de Fisica del Cosmos de Arag\'on, Plaza San Juan 1, Planta 2, Teruel, E44001, Spain
\next
European Space Astronomy Centre, P.O. Box 78, 28692 Villanueva de la Ca\~{n}ada, Madrid, Spain}
\mainauthor{luca.izzo@gmail.com}
\maketitle

\begin{abstract}%
We discuss the nebular phase emission during the 2011 outburst of the recurrent nova T Pyxidis and present preliminary results on the analysis of the line profiles. We also present some discussions about the binary system configurations and the X-ray emission, showing that the white dwarf mass should be larger than 0.8 M$\odot$.
\end{abstract}

\keywords{Spectroscopy - X-rays - individual: T Pyx}

\begin{multicols}{2}

\paragraph{1. Introduction}

The 2011 outburst of the recurrent nova T Pyx shed more light on some interesting characteristics of this peculiar nova.
The optical evolution was not very different from the last (1966) outburst, showing the 8 days 'shoulder' before the final rise to the maximum luminosity, which was reached $\sim$ 30 days from the initial outburst. The distance of T Pyx was determined from light echoes to be 4.8 $\pm$ 0.5 kpc \cite{Sokoloski}, which is marginally consistent within the errors with the value of 3.5 $\pm$ 0.4, estimated from the application of the maximum magnitude versus rate of decline (MMRD) relation \cite{DellaValle1998} to the light curve of the 1966 outburst \cite{Selvelli}. The presence of a clumpy ring with an inclination of 30-40 degrees, embedded within a spherical shell, was also recently reported using Hubble Space Telescope observations during the nebular phase \cite{Sokoloski}. A variation of the orbital period after the 2011 outburst was detected, indicating an ejected mass in the outburst of $3 \times 10^{-5}$ M$\odot$ \cite{Patterson2013}. Also, the orbital period was found to have increased during the long quiescence period between the last two outbursts of T Pyx. 
Moreover, with respect to the 1966 epoch, when no advanced X-ray and radio detectors were available, we have now a good knowledge of the emission of T Pyx at these frequencies during its outburst \cite{Nelson}. 

In the previous work \cite{Izzo}, we have presented the early spectral evolution, up to two months after the initial outburst. We have shown the hybrid spectral transition He/N - Fe II - He/N and initial ejecta velocities. In this work, we present the optical observations of the nebular phase, which was marked by the onset of the supersoft X-ray source (SSS) emission which happened during the seasonal gap \cite{ATEL}. The optical observations presented here were obtained with the high-resolution spectrograph SARG, mounted on the 3.5m telescope Nazionale Galileo (TNG) located at La Palma. Spectra with power resolution of R = 29000 were obtained but the complete sample, which includes also observations obtained at the Very Large Telescope, will be presented in Ederoclite et al. in preparation. Then we briefly discuss the X-ray emission and finally we provide an updated scheme of the binary system configuration in T Pyx, taking into account the recent results.

\paragraph{2. The nebular phase of T Pyx}

The nebular phase in novae is easily recognized by the absence of absorption lines and the presence of forbidden emission lines, e.g. [O\ III] and [N\ II], in the optical spectra, similarly to planetary nebula cases. Indeed, these emission lines are due to photo-ionization of the ejecta by the central hot white dwarf (WD). At the same time, forbidden transitions originate mainly in rarified environments, so that any structure observed in corresponding line profiles is mainly due to the distribution of the material and not to self-absorption mechanisms. The analysis of the [O\ III] 4959-5007 and [N\ II] 5755 $\lambda\lambda$ profiles can consequently provide hints on the geometry of the ejecta.

These line profiles are well described by a large flat top base ($v_{exp,1} \approx 1700$ km/s) overlapped by a saddle-shaped and brighter emission, with $v_{exp,2} \approx 600$ km/s, see Fig. \ref{fig:no1} for the case of the [O\ III] 5007 $\lambda$. The high resolution allows us to detect a castellated structure on the top of the profiles which is due to the presence of a clumpy ejecta structure. The saddle-shaped configuration suggests a non-spherical geometry of the ejecta, as generally discussed in \cite{Payne} and further developed in several publications. In order to estimate the asphericity from forbidden lines we have used a method first presented in \cite{Bappu}. An analytic expression for the line profiles $I(l)$ from an anisotropic distribution of the nova ejecta was provided:
\end{multicols}

\begin{eqnarray}\label{eq:no1}
I(l) = K \left(\frac{c}{v_0}\right)^{\alpha +1} \times \nonumber \\
\left[\frac{l_2^{\alpha-2}-l_1^{\alpha-2}}{\alpha-2} l^2 (1-3 cos^2 i) +  \frac{l_2^{\alpha}-l_1^{\alpha}}{\alpha} (1- sin^2 i)\right],  & l \leq l_1; \nonumber \\
\left[\frac{l_2^{\alpha-2}-l^{\alpha-2}}{\alpha-2} l^2 (1-3 cos^2 i) +  \frac{l_2^{\alpha}-l^{\alpha}}{\alpha} (1- sin^2 i)\right],  &  l \geq l_1;  \nonumber \\
\end{eqnarray}

\begin{multicols}{2}

where $c$ is the speed of light, $v_0$ is the velocity at the inner radius of the expanding shell (we have assumed the maximum velocity observed for the bright profile, $v_0 \approx 600$ km/s), $\alpha$ describes the variation of the shell's velocity with the distance form the WD, and $l = (\lambda-\lambda_0)/\lambda_0 = v/c$ is our variable. The quantity $l_2 = v_2/c$ corresponds to the zero intensity along the wing of the line profile. A grid-search algorithm was developed to find the best fit for the inclination angle $i$ that the binary orbital plane forms with the line of sight. For all the three lines, we find a large angle $i \approx 60 \pm 10$ degree, see also Tab. \ref{tab:no1}, which contrasts some estimation of the $i$ given before the 2011 outburst \cite{Uthas}, and suggests a larger value for the binary inclination angle, see also \cite{Sokoloski}. We note that this approach does not take into account the presence of a clumpy structure. A more detailed approach is needed, and will be presented elsewhere.

 \begin{table*}
\begin{center}
\begin{tabular}{lcccc}
\hline\hline
Ident. wavelength ( $\AA$) & $i$ (degree)      &  $v_1$ (km/s)  & $\alpha$ & K  (erg/cm$^2$/s/Ang)  \\
\hline
 4959 & 59 $\pm$ 10 & 465 $\pm$ 6 & -2.2 $\pm$ 0.3 & (3.7 $\pm$ 0.1) $\times$ 10$^{-15}$ \\
 5007 & 59 $\pm$ 10 & 477 $\pm$ 6 & -2.3 $\pm$ 0.3 & (1.09 $\pm$ 0.03) $\times$ 10$^{-14}$ \\
 5755 & 60 $\pm$ 10 & 444 $\pm$ 6 & -2.8 $\pm$ 0.3 & (6.8 $\pm$ 0.5) $\times$ 10$^{-16}$ \\
\hline
\end{tabular}
\caption{Fit results of the line profiles of [O\ III] 4959, 5007 and [N\ II] 5755, using the approach developed in \cite{Bappu}.}
\label{tab:no1}
\end{center}
\end{table*}

\begin{myfigure}
\centerline{\resizebox{80mm}{!}{\includegraphics{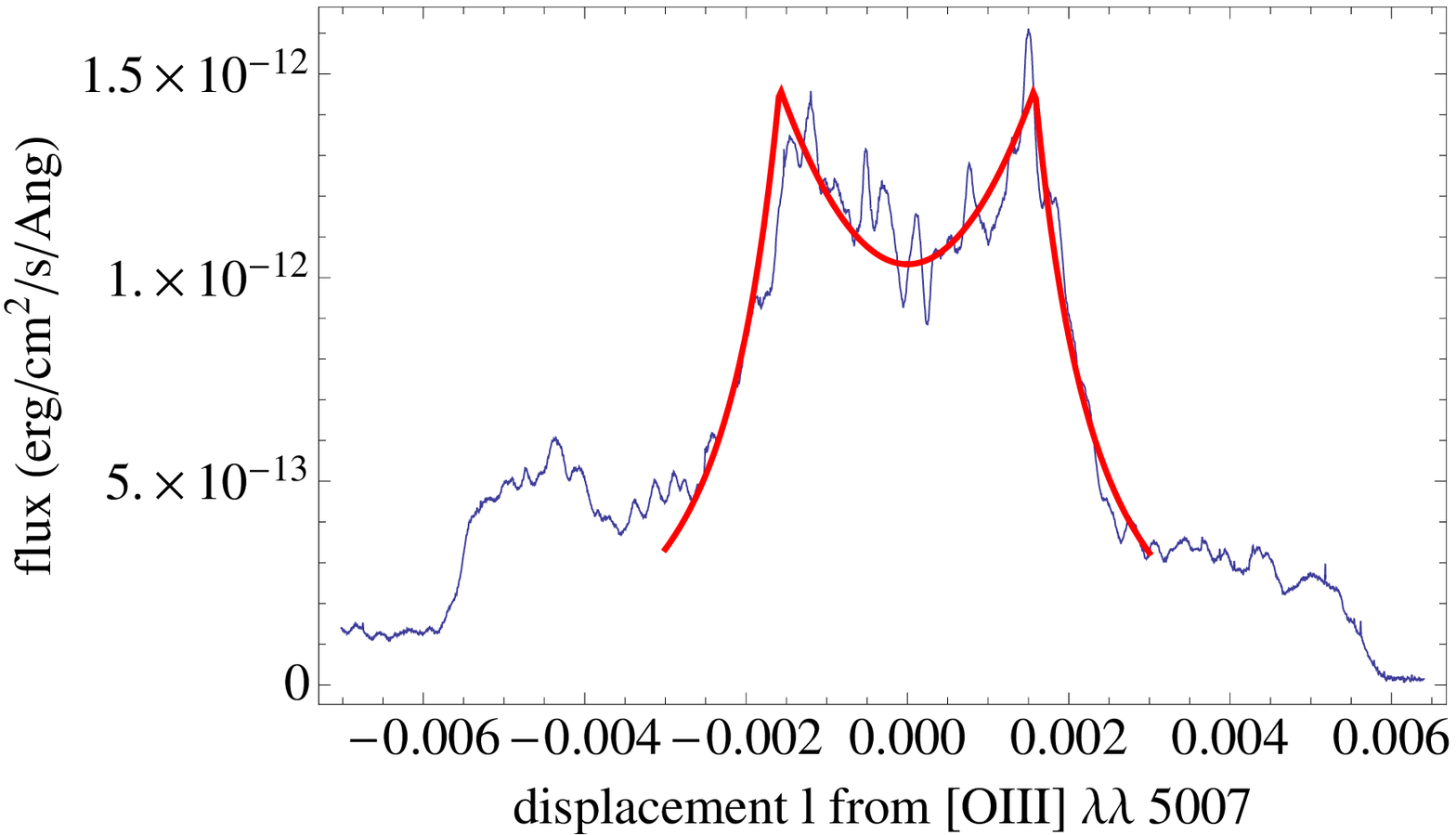}}}
\caption{The saddle-shaped profile in the [O\ III] 5007 line and the best fit (red line) obtained with the method described in the text.}
\label{fig:no1}
\end{myfigure}

\paragraph{3. The configuration of the binary system}

It is clear from the nebular spectra that the emission in T Pyx is well described by quasi-spherical ejecta, which give rise to the larger flat top profile, and an equatorial ring of material, which is slower and gives rise to the saddle-shaped profiles. The origin of this equatorial ring is unknown: it is possible that it formed during the interaction of the main ejecta with pre-existing gas, which is equatorially distributed around the binary system (an hypothesis presented in \cite{Williams2012}); another possibility can be that the early interaction of the ejecta with the secondary star could influence the subsequent expansion \cite{Hutchings}, but other mechanisms can also work. In all cases, it is necessary to shed light on the observed asphericity in T Pyx, and the first ingredient we need is the mass of the WD.

X-ray observations of the SSS phase can in principle provide this physical quantity. The SSS emission is caused by stable hydrogen burning in the matter that was not ejected during the nova outburst. This burning matter forms an envelope around the WD and can be observed once the ejected material becomes optically thin to soft X-rays. The duration of the SSS phase is inversely related to the WD mass \cite{Starrfield1991}, since more massive WDs need to accrete less matter to trigger an outburst. Given its recurrent nature, T Pyx is expected to contain a massive WD.

The SSS phase of T Pyx was first detected about 90 days after the outburst \cite{ATEL}, when an increase in soft X-rays was observed by the  Swift-XRT \cite{Burrows}. T Pyx remained in this state for $\approx$ 200 days, although around day 190 the luminosity started to decline. We do not enter here into a discussion on the details of this emission, which will be described in a forthcoming work. What is interesting in the context of this work is the relatively short duration of the SSS phase, which suggest a large WD mass. 

Observations obtained before 2011 reported a WD mass of 0.7 M$\odot$ and a mass ratio of $q = 0.2$ \cite{Uthas}, which implies a secondary star of 0.14 M$\odot$. The orbital period was observed with high accuracy to be $P_{orb} = 1.8295$ hrs. With some assumption on the type of the companion star, we can infer the mass of the secondary which overfills its Roche lobe: $\Delta M_2 = \rho_2 \frac{4}{3} \pi (\Delta R_2)^3$, where $\rho_2$ is the density and $\Delta R_2$ the dimensions of the overfilling envelope of the secondary. From the $q$ value and the absence of signatures of the secondary in optical spectra during the quiescent phase \cite{Selvelli}, we can assume that it is a low-mass M-type star. Its density can be well approximated using the Eggleton formula $\rho_2 = 107 P_{orb}(hrs)^{-2}$ g cm$^{-3}$ \cite{Eggleton}, while its radius can be derived from the Patterson et al. formula $R_2 = 0.62 M_2^{0.61}$ R$\odot$ \cite{Patterson2005}. The Roche lobe size of the secondary is related to the $q$ parameter through $L_2 = 1.631 q^{1/3} M_1^{1/3} P_{orb}(hrs)^{2/3}$ 10$^{10}$ cm \cite{Paczynski}.

The result is that the envelope overfilling the secondary Roche lobe is related to the WD mass. From the recent estimate of the mass ejecta after the 2011 outburst ($3 \times 10^{-5} M\odot$, \cite{Patterson2013}), and assuming that every observed T Pyx outbursts ejected similar masses, we obtain a total mass ejected from T Pyx of about $2 \times 10^{-4}$ M$\odot$. If this mass was accreted onto the WD from the secondary we can obtain a lower limit on the WD mass from the above formula. This results in $M_{WD} \geq 0.8$ M$\odot$.

\paragraph{4. Final Considerations}

\begin{figure*}
\centering
\includegraphics[width=0.9\hsize,clip]{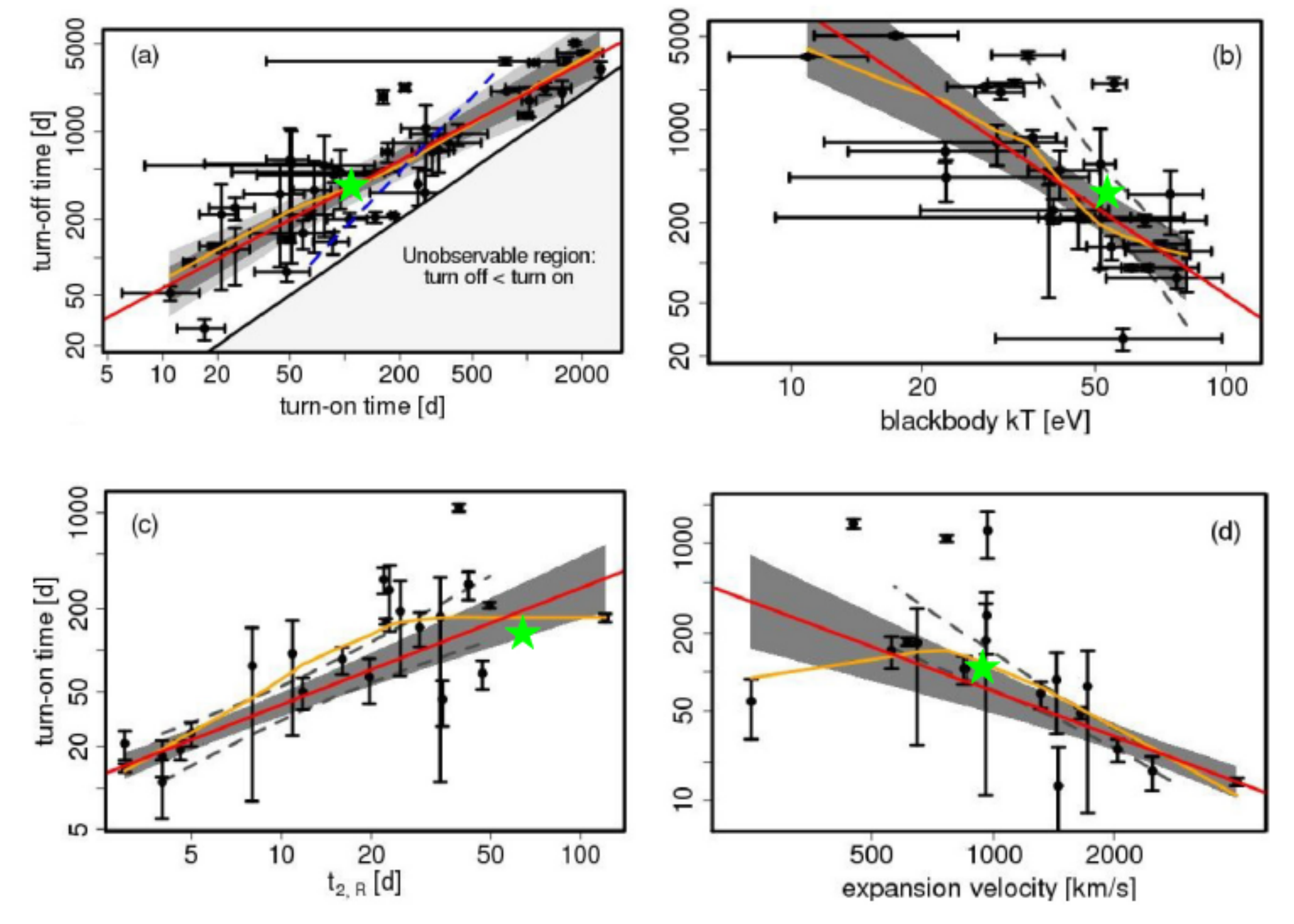}\\
\caption{The location of the T Pyx parameters (green stars) with respect to the M\,31 nova correlations presented in \cite{Henze}: (1) the t$_{on}$-t$_{off}$ correlation, for T Pyx measured from the time of maximum brightness; (2) the kT-t$_{off}$ relation, with kT given by the average value kT=55.7 eV; (3) the relation between t$_{on}$ and t$_2$ and (4) the correlation between t$_{on}$ and the expansion velocity as measured from the $H\alpha$ emission line.}\label{fig:no2}
\end{figure*}

Our last result requires some important remarks. First, it was proposed that the mass ejected in T Pyx is larger than the one accreted \cite{Selvelli}.  We do not consider this possibility in this work, i.e., that all the accreted mass results to be the ejected one. We know that this is not exactly true, since the SSS phase is due to the burning of accreted hydrogen which is not ejected in the nova outburst. The mass of this burning envelope can be estimated assuming that the energy source for the SSS phase is the gravitational contraction \cite{Krautter}, but this contribution can be neglected. Second, there is the possibility that a fraction of the matter overfilling the secondary's Roche lobe would flow out from the external Lagrangian point $L_2$, leading to the formation of an asymmetric structure around the binary system. Such a mechanism would operate only if the primary Roche lobe has been filled by accreted gas, leading to the detection of transient heavy elements in absorption (THEA) during the early phases of the outburst \cite{Williams2008}, whose presence, in this case, is not certain. Moreover, in tight systems where the secondary is less massive than the WD ($q < 1$) also centrifugal forces can contribute to such an outflow. There is again observational evidence that T Pyx shares many properties common to other novae. As an example of this, we compare T Pyx to the unbiased sample of M31 novae \cite{Henze}, in particular with respect to the correlations between some optical and X-ray observables: the blackbody temperature of the SSS emitter, the turn-on and turn-off time of the SSS emission, the expansion velocity of the ejecta and the $t_2$ determined from the R-band light curve \cite{DellavalleLivio}, see Fig. \ref{fig:no2}. We see that T Pyx lies in the same region occupied by classical novae observed in M31.  All these results need to be figured in a unique picture for T Pyx and can shed more light on CV systems with a mass ratio of less than unity. Further results and conclusions will given in Izzo et al. (2015) in preparation.  

\thanks
We are very grateful to R. Williams, S. N. Shore, E. Mason, D. de Martino and R. Gilmozzi for precious discussions. M.H. acknowledges support from an ESA fellowship.

\bigskip
\bigskip
\noindent {\bf DISCUSSION}

\bigskip
\noindent {\bf CHRISTIAN KNIGGE's Comment:} The system parameter we estimated in Uthas et al. 2011 should be used with caution. They are based on a radial velocity study, and parameters resulting from such studies are known to be biased quite often.

\bigskip
\noindent {\bf CHRISTIAN KNIGGE's Question:} You consider $R_2$ and $L_2$ differing by $\approx$ 20$\%$. Are you really saying this could be the physical configuration, i.e. that the donor is overflowing the Roche lobe by this amount ?

\bigskip
\noindent {\bf LUCA IZZO:} The results that I've presented comes out naturally from formulae and physical considerations based on fixed binary parameters, as $M_{WD} = 0.7$ M$\odot$, $q = 0.2$, $P_{orb} = 1.8295$ hrs, and considering the donor as a low mass M-type star. Following a work by Patterson et al., we obtain the $R_2$ value, while from Paczynski the value of $L_2$. Their interpretation depends obviously on a binary parameter as $q$ presented in the Uthas et al. work. Anyway, other works consider that value of $q$ as the actual one working in T Pyx case. 

\end{multicols}
\end{document}